# Quantum information density scaling and qubit operation time constraints of CMOS silicon based quantum computer architectures

Davide Rotta[1,2], Fabio Sebastiano[3], Edoardo Charbon[3], and Enrico Prati[1]

[1]*Istituto di Fotonica e Nanotecnologie,*
*Piazza Leonardo da Vinci 32 – I, 20133 Milano*
[2]*TeCIP Institute - InPhoTeC, Scuola Superiore Sant'Anna,*
*Via G. Moruzzi, 1 – 56124 Pisa – Italy*
[3]*Delft University of Technology, Faculty of Electrical Engineering,*
*Mekelweg 4, Delft, The Netherlands*

## Abstract

Even the quantum simulation of an apparently simple molecule such as $Fe_2S_2$ requires a considerable number of qubits of the order of $10^6$, while more complex molecules such as alanine ($C_3H_7NO_2$) require about a hundred times more [0]. In order to assess such a multimillion scale of identical qubits and control lines, the silicon platform seems to be one of the most indicated routes as it naturally provides, together with qubit functionalities, the capability of nanometric, serial and industrial quality fabrication. The scaling trend of microelectronic devices predicting that computing power would double every two years, known as Moore's law [1], according to the new slope set after the 32 nm node of 2009, suggests that the technology roadmap will achieve the 3-nm manufacturability limit proposed by Kelly [2, 3] around 2020. Today, circuital quantum information processing architectures are predicted to take advantage from the scalability ensured by silicon technology. However, the maximum amount of quantum information per unit surface that can be stored in silicon based qubits and the consequent space constraints on qubit operations have never been addressed so far. This represents one of the key parameters towards the implementation of quantum error correction for fault-tolerant quantum information processing and its dependence on the features of the technology node. The maximum quantum information per unit surface virtually storable and controllable in the compact exchange-only silicon double quantum dot qubit architecture is expressed as a function of the CMOS technology node, so the size scale optimizing both physical qubit operation time and quantum error correction requirements is assessed by reviewing the physical and technological constraints. According to the requirements imposed by the quantum error correction method and the constraints given by the typical strength of the exchange coupling, we determine the workable operation frequency range of a silicon CMOS quantum processor to be within 1 and 100 GHz. Such constraint limits the feasibility of fault-tolerant quantum information processing with CMOS technology only to the most advanced nodes. The compatibility with classical CMOS control circuitry is discussed, focusing on the cryogenic CMOS operation required to bring the classical controller as close as possible to the quantum processor and to enable interfacing thousands of qubits on the same chip via time-division, frequency-division, and space-division multiplexing. The operation time range prospected for cryogenic control electronics is found to be compatible with the operation time expected for qubits. By combining the forecast of the development of scaled technology nodes with operation time and classical circuitry constraints, we derive a maximum quantum information density for logical qubits of 2.8 and 4 Mqb/cm$^2$ for the 10-nm and 7-nm technology nodes respectively for the Steane code. The density is one and two orders of magnitude less for surface codes and for concatenated codes respectively. Such values provide a benchmark for the development of fault tolerant quantum algorithms by circuital quantum information based on silicon platforms and a guideline for other technologies in general.

The efforts towards large scale quantum information processing (QIP) for practical applications have been boosted after two key milestones: the advent of quantum algorithms [4, 5] and the invention of quantum error correction codes [6–10], including the implementation of error correction in a fault-tolerant manner with concatenated [11–14] and topological codes [15–18]. Such methods call for a very large number of physical qubits, a requirement that has triggered the quest for a silicon platforms suitable for manufacturing large arrays of devices. In addition, the mature silicon nanoelectronics platform may play a role in providing the extensive classical circuitry that can meet the speed and power specifications required to manipulate and readout such large qubit arrays. Finally, the protection of quantum coherence when quantum states are shuttled across multiple interdevice distances represents a difficulty which may be overcome by exploiting the nanometric scale size of the most aggressive technology nodes. To this extent, the maximum amount of quantum information density per unit surface is not a mere curiosity [19, 20], but it represents a key technological boundary since the size of quantum computer chips will be reasonably limited to few $mm^2$ for packaging and refrigeration considerations.

Silicon platforms for quantum computing hold the promise to achieve high scalability, testability, and reliability levels [21, 22], however a straightforward integration of multiple silicon spin qubits with CMOS control electronics is not straighforward and a complete architecture for such an interface still lacks. In particolar, the quantification of the scalability potential for this interface, its node-dependent performance and the impact on the implementation of algorithms and quantum error correction (QEC) pose precondition issues that have never been addressed systematically. Cryogenic Complementary Metal-Oxide-Semiconductor (CMOS) circuits for readout of quantum states in semiconductor quantum dots (QD) have been developed in the past [23–26]. CMOS based quantum dots have been demonstrated [27], while small arrays of qubits have been employed for elementary operations in GaAs [32–34] and non-CMOS silicon quantum dots [35–37]. By reviewing the state-of.the-art of circuital QIP and of its control electronics in silicon, we analyze the constraints binding the assessment of a fully CMOS approach that combines classical electronics with quantum circuits on the same substrate operating at cryogenic temperatures. To achieve such a goal, for technical reasons discussed later, we select the exchange-only silicon quantum dot qubit [38, 39] based on three electrons in a double quantum dot. In fact, this qubit is the most compact to be implemented with all-electrical control in semiconductor technology and despite the important development required to scale up this architecture, its features make it worthwhile to conduct a careful consideration of its advantages and drawbacks.. Furthermore, the discussion can be relatively easily adapted to single-triplet (S-T) spin qubits in double quantum dots, at the cost of increasing the complexity of the system by adding magnetic field gradients. Let us first define a few elementary building blocks for the proposed architecture to compose scalable quantum circuits with classical readout and control. Inspired by the International Technology Roadmap of Semiconductors (ITRS), we determine the technology-dependent area of the key-devices generating universal quantum circuits. The area of logical qubits is consequently calculated for two QEC architectures, namely Steane code [7,1,3] and surface code, as to obtain the scaling law of quantum information density as a function of the technology node for both. The implications on the realization of scalable quantum algorithms as well as the integration with classical control circuitry is then discussed with a special attention to the matching of the operation time of the qubits with the control electronics. A general layered architecture includes physical, virtual, quantum error correction, logical and application layers respectively [0]. A silicon CMOS substrate may provide a physical realization of the physical layer of the qubits, the virtual layer by integration of classical control circuits for dynamical decoupling compensation sequences and readout, and the quantum error correction layer. Cryogenic operation of the classical control circuitry is required to enable co-integration with the quantum processor, thus enabling a simplified interconnection scheme to a number of qubits large enough for fault-tolerant quantum processing. Such cryogenic time-division, frequency-division, and space-division multiplexing can significantly reduce the number of interconnects and power dissipation, by operating in a frequency range between 1 and 10 GHz, which is compatible with the operation time range imposed by physical and architectural constraints of logical qubits based on exchange-only three spin qubits in silicon double quantum dots, the latter being around 1-100 GHz.

# I. GEOMETRICAL CONSTRAINTS OF THE HYBRID QUBIT ARCHITECTURE

Logical qubits operated by a substrate-independent application layer are based on the fault-tolerant quantum error correction circuit implementing classical operations for the manipulation of virtual qubits and virtual gates. In turn, the virtual layer is based on the measurement and manipulation of physical qubits, implemented throughout our review by electron (hole) spin in CMOS silicon quantum dots. In this section we discuss how the virtual layer is assessed from the physical layout by keeping the space resources at the maximum compactness.

Several virtual qubits have been proposed for QIP in Si [115]. In single spin physical qubits, the two spin states of an electron in a quantum dot are Zeeman-split by an external magnetic field and manipulated by means of microwave pulses [33, 40]. A singlet-triplet virtual qubit architecture is based on two electron spins in a double quantum dot and it does not require microwave pulses [32, 35]. Fast operation is achieved through electrostatic control, provided that a strong magnetic field gradient is built across the double quantum dot. Both architectures were implemented in silicon, demonstrating coherent qubit rotations at GHz frequencies, which is much faster than the coherence time predicted for such QD systems ($T_2^* \sim \mu$s) [38]. The integration of micron-sized resonant microwave antennas and micromagnets for every single qubit may be a major issue against the large scale integration of single spin and singlet-triplet qubits respectively, as state of the art microwave line integration for qubit control ranges around 100 microns length scale of minimum feature size [41]. Inhomogeneity in the magnetic field gradient and microwave interaction with nearest qubits may introduce strong variability in the qubit functional properties and consequently high error rates at the virtual layer. On the contrary, an architecture featuring all-electrical manipulation could take the best of the CMOS technology in terms of scalability and natural integration of the classical control circuitry. This requirement is met by using three electron spin qubits, whose dynamics is governed only by their exchange interaction. Originally, an architecture of this kind was proposed in systems of triple quantum dots, where the complete control of qubit states is obtained by tuning the inter-dot exchange couplings [34, 42]. Later, a more compact version of the exchange-only triple electron spin virtual qubits has been proposed [38, 43] and implemented [37] by using only two quantum dots, one of which is doubly occupied. Both architecture are deeply reviewed in Ref [115] In contrast to previous architectures, only a small magnetic field is required during the virtual qubit initialization, provided externally [38, 42]. Therefore, one may take this qubit as the most compact one with all-electrical control.

The exchange-only DQD qubit, sometimes referred to as a hybrid qubit for its intermediate nature between charge and spin qubits, employs three electron spins in a semiconductor DQD.

The logical basis is defined in the spin subspace with total spin $S = 1/2$ and vertical component $S_z = -1/2$. The logical states employed in the computation are:

$$|0\rangle \equiv |S\rangle|\downarrow\rangle, \quad |1\rangle \equiv \sqrt{\frac{1}{3}}|T_0\rangle|\downarrow\rangle - \sqrt{\frac{2}{3}}|T_-\rangle|\uparrow\rangle \tag{1}$$

where $|S\rangle$ and $|T\rangle$ denote the singlet and triplet states of the doubly occupied dot, whereas $|\uparrow\rangle$ and $|\downarrow\rangle$ indicate the spin orientation of the single electron in the second dot. Such a virtual qubit is set in the $|0\rangle$ state by operations at the physical layer by firstly polarizing the uncoupled spin in a magnetic field that is briefly switched on. Although no magnetic field has been used in single qubit experiments [37], this is required when dealing with more qubits in order to set a common reference for the spin orientation in different qubits. At operating temperatures of about 100 mK, a magnetic field intensity of 1.5 T is sufficient to impose the orientation of a single spin by means of spin-dependent tunneling from a reservoir to the QD [55]. We also note that, although not necessary for the manipulation, a small magnetic field could even be beneficial in mitigating the effect of the low-frequency magnetic field fluctuations [116]. The time load for this operation corresponds to the electron tunneling time, which can be set exceedingly small by lowering the tunnel barrier. This technique leads to an initialization of the singly occupied dot with fidelity currently of 95%, which is limited by thermal noise [36, 77]. The doubly occupied dot is then initialized by driving the system in a configuration, where singlet is the ground state. The manipulation of the virtual qubit is performed by tuning the inter-dot effective exchange interaction. The effective Hamiltonian driving the system can be written in term of spin-spin interactions only [44]. Such interactions are finely tuned by controlling the quantum dot potential and the inter-dot electrostatic barrier, allowing fast and all-electrical manipulation of the qubit state. The first experimental work demonstrated coherent qubit rotations at GHz frequencies with

fidelity of about 90% [37], albeit much better performances are theoretically predicted with optimized fabrication and pulse sequences [43, 116, 117]. The final state is read out by means of charge sensing after collapsing the DQD system in either the (2,1) or the (1,2) charge state, corresponding to the $|0\rangle$ and the $|1\rangle$ logic state respectively. Two-qubit gates could be performed similarly as in other qubit architectures by exploiting either the capacitive or the exchange coupling between two adjacent qubits [107-109, 116] and so providing a universal set of virtual gates [39,45]. The exchange-only qubit has been realized in hepitaxial Si-SiGe heterostructures and Rabi oscillations have been observed [37], with a $T_2^*$ = 2 ns at nominal electron temperature of 150 mK and controlled at picosecond timescale. In principle, a complementary system based on holes [46] instead of electrons is also possible by using CMOS technology [47,48]. Holes in silicon carry the potential advantages of a considerably smaller hyperfine coupling with nuclei that cause decoherence [49, 50], and lack of the valley degeneracy that both causes interference phenomena [51] and interleaved complicated valley sub-orbitals as happens for low electron filling [52]. Therefore, most of the following discussion holds for both electron and hole double quantum dots. On the contrary, a different reasoning applies to donor impurities in silicon [53, 54]. In brief, single electron spin donor qubits exhibit very long coherence time of the order of seconds [29, 55], and the effectiveness of atomic resolution lithography based on scanning tunneling microscope [56] is progressively approaching serial implantation [57]. The less accurate single ion implantation method (SII) [58] could achieve sufficient precision for some architecture such as a surface code implementation based on a two-dimensional array of distant donors [59] which tolerates deviation of up to 11 nm from the ideal lattice position. However, the complexity of the serial design of devices involving either individual donors with single spins for qubits with microwave control, or pair of donors with two or three electrons bound to donors controlled by gates that depends on a relatively high number of currently unaddressed assumptions including yield of implantation and activation of all the donor sites, control of inter-donor distance at single lattice precision, global or individual microwave control, singlet-triplet and exchange-only three spin qubits, CMOS mask design on top of silicon overgrowth on the donor layer. The surface code implementation proposal based on a spin-probe controlled two-dimensional array of donors [59] considers a distance of about 400 nm between neighboring donors, which represents an intermediate scale between quantum dots and superconductor qubits. Regardless of the chosen physical implementation, besides the desirable improvements concerning the single qubit specifications to achieve fault-tolerant fidelity for one- and two-qubit gates, the practical demonstration of two-qubit devices and more complex circuits involves additional issues [115-116]: the need for long range quantum communication, quantum error correction and the demanding requirements related to the massive integration of classical control electronics at the quantum chip level. In the following we focus on the double quantum dot design for hybrid qubits, which could also be naturally adapted to charge qubits [109] and singlet-triplet architectures [43] if the issue of adding a local magnetic field gradient can be addressed with no cost in terms of additional space requirements.

As one is interested in the large-scale fabrication of quantum circuits based on a silicon platform, the CMOS implementation of the hybrid qubit architecture is considered, towards the identification of the scaling law of computational power per unit surface area as a function of the technology node. Universal QIP can be addressed by different approaches. First, we consider the universal set of single virtual qubit rotations and virtual CNOT logic ports [39] for a Steane code [7,1,3], so we adopt the definition of the three physical building blocks [60], namely a data qubit (D) capable of one- and two-qubit logic gates and two types of communication qubits i.e. the chain module (C) and the T module (T). A schematic representation of these devices is reported in Figure 1.

The module D incorporates two qubits (i.e. four quantum dots) and the corresponding individual electronic reservoir and Single Electron Transistor (SET) for independent initialization and readout. To assess the 1-qubit and 2-qubit gates mechanisms, as well as measurement readout at the physical layer, the system is equipped with a reservoir consisting of high electron density region with a quasi-continuous density of states that is controlled by an accumulation gate to provide electrons required for qubit manipulation during the initialization procedure. The SET is used as a single charge sensor to monitor the qubit charge state during readout. The coherent transfer of quantum information between distant qubits is a very challenging task which could in principle be assessed through the sequential repetition of SWAP logic gates between adjacent qubits, as proposed in Ref. [60]. The module C is specifically designed for quantum communication following this procedure, with no need for initialization and readout. To this purpose, the use of different techniques, such as the coherent transfer by adiabatic passage and teleportation, will be also considered later, as well as the impact of electron loss during the qubit transportation. Finally, the module T is a modified version of the latter to connect perpendicular quantum communication lines and create two-dimensional arrays of qubits. Each device is developed within the same design rules for a given technology node and instantiated as a conventional component in the design of large arrays of identical qubits independently accessible by classical electronics (for more extensive review on the interface between the physical qubits and the classical electronic layer see Ref. [112]). The physical size of the three building blocks constituting the virtual layer is then calculated as a function of the main critical sizes such as the pitch of metal interconnections and the width of Si islands. For example, module C height is constrained by the silicon wire width and by the four vias on each side. The sizes of the other devices are determined analogously in terms of the critical pitches. The expressions reported in Table I are defined only by the device layout, while being totally independent of the technology node. $\Delta_G$ and $\Delta_{IC}$ are the contacted gate pitch and the interconnect pitch respectively, $w_{Si}$ is the width of the silicon wires hosting the dots, $l_{HDD}$ is the length of the highly doped drain and $l_{BU}$ is an undoped silicon buffer to avoid unintentional doping of the device active region.

As an alternative option, we examine the implementation of surface codes based on nearest neighbor qubit lattices, and having marked differences from the device layout envisaged for the Steane code. Surface codes appear to be promising candidates for the implementation of fault tolerant quantum circuits with error thresholds in the $10^{-2}$ range. Notably, a possible layout was proposed for the implementation of both QEC and leakage correction protocols by surface codes based on singlet-triplet qubits realized in semiconductor double quantum dots [87]. By exploiting the formal and geometrical analogy between the singlet-triplet qubit and the hybrid qubit here discussed (both based on semiconductor double QDs hosting electron spins) it is possible to evaluate the size of the consequent logical qubit for surface codes [18] as the two cases may be treated with no fundamental difference [105], with the exception of the highest operation speed for the hybrid qubit, which is the case considered in this review.

A layout for the implementation of the surface code on a nearest neighbor lattice of physical qubits is depicted in Figure 2 (fourth column), in analogy with [87]. The asymmetry of the hybrid qubit does not constitute a difficulty as the four hybrid qubits can be arranged in an alternated configuration block (top of fourth column of Figure 2). The surface code consists of a 2-dimensional lattice of data (red) and ancillae (green and yellow) qubits to implement Z- and X-stabilizers for quantum error correction by classical operations. The physical qubit area is estimated accordingly as the area of such block divided by 4 i.e. $(8\ \Delta_G)^2 / 4$. Nearest neighbor connections impose stringent proximity of the physical qubits, which from one side lead to a minimum footprint, but on the other raises potential difficulties in their control to address the physical operations to achieve an operating virtual layer. The SET-based charge sensors proposed for the Steane code may be replaced by rf-reflectometry sensors fully integrated within the control gates [71,72]. It is worth mentioning that such compact layout comes at the expense of considerably higher complexity at the classical control layer, discussed later. The classical circuitry architecture is supposed to be able to deliver rf pulses of arbitrary shapes for qubit manipulation as well as readout, dealing with serious issues of high density routing and cross-talk behavior.

While in the first experimental works on single qubit devices this has not been a blocking point, the complexity increases in a large scale computer. The replication of the same wiring strategy is not a viable solution for two reasons: on the one hand, the extremely dense topology of physical qubit arrays generates serious routing problems for the interconnections on the quantum chip; on the other hand, interfacing billions of control lines with classical electronics on a different substrate or package is not realistic with current and foreseeable back-end technologies.

It is widespread opinion that the classical electronics will be split into several stages between the quantum chip and room temperature electronics (see Fig. 5a) to optimize the system thermal management. A classical integrated circuit directly attached on the quantum chip by conventional flip-chip technology may provide the low-level interface with the quantum layer. An interposer with superconducting through-silicon vias could be effective in limiting the thermal load arising from classical electronics. Nonetheless, the partial integration of control electronics within the quantum chip level is an option that should be considered to alleviate routing issues at the quantum chip level, to make the interface with higher level classical electronics easier and to improve its performance by reducing the distance from the qubits [112].

Starting from both the three building blocks D, C and T for the Steane code [7,1,3], and the double dot qubit for the surface code respectively, logical qubits and circuits for implementation of quantum algorithms can be designed. Figure 2 shows the symbolic notation, functional and physical representations of QIP at different scales of integration of the two architectures, from the virtual qubit to quantum error correction circuit, to logical gate and application layer fault tolerant quantum chip scale respectively.

By following the typical arrangements of virtual qubits proposed for scaling to a logical qubit and as a further step to H-tree structures for concatenated codes [84], the first level of integration (second row in Figure 2) for the Steane code is achieved by connecting a relatively small number (~ 20) of virtual qubits (device D) by means of quantum channels (modules C and T) to create the physical background for fault-tolerant quantum computing. The corresponding physical device is sketched at the third column by means of green blocks (communication modules namely C and T) and red boxes (data qubit D) following the symbolic representation defined in Figure 1. This structure defines the smallest system of virtual qubits connected to a line of bidirectional quantum communication with the scope to store quantum information and correct potential errors. A number of seven virtual qubits is needed to define a logical qubit according to the Steane's code, and about twenty virtual qubits suffice to correct potential X and Z errors on a logical qubit according to most quantum codes e.g. the Shor's and the Steane's ones [61]. As long as the error rate is lower than the code error threshold, arbitrary errors occurring at the individual physical qubits can be detected and corrected in the framework of a logical qubit by employing few additional qubits as ancillae. We note that previous theoretical works identified several strategies to obtain gate fidelity approaching 99.999 % [43,60]. This may be good enough to enable QEC by means of the Steane code, which could deal with an error threshold in the range $10^{-6}$ – $10^{-4}$ depending on the geometry [61,112]. So, if the error rate is a factor of $x$ better than threshold, encoding yields a final error rate approximately a factor of $x^2$ below threshold. Further improvement can be achieved similarly by concatenation [61]. At any rate, we remark that only a deep understanding of the noise model applicable to the hybrid qubit will definitely set the relevant specifications for gate fidelity to be compatible with a specific QEC scheme [60,61]. Such logical qubit therefore enables QEC and it is the functional building block for error-correcting quantum memory and fault-tolerant QIP. Differently, the implementation of surface code in the fourth column of Figure 2 exploits the physical qubits in a 2D nearest neighbor array so that the data virtual qubits (red ovals) are interleaved with ancillae for the X (yellow) and Z (green) stabilizer measurements.

The next length scale (third row) requires interconnections between multiple logical qubits to perform small-scale algorithms (in the first column, the Quantum Fourier Transform is given as an example) in a fault-tolerant manner. At such level, in order to estimate the effective area occupied by a single logical qubit by including the space needed to connect the qubits in some arrangements, for example by an H-tree structure [84], it is useful for the Steane code to *conventionally* define the logical qubyte or quantum byte by 8 connected logical qubits, as a reminiscence of classical information processing. The qubyte allows calculating the minimum effective area of a single logical qubit by dividing by 8 its area. Such operation is not needed for the surface code architecture, which employs contiguous logical qubits. An example of operation carried at this level is provided by Quantum Fourier Transform, a key block of more complex algorithms like Shor's factoring. Finally, many logical qubit blocks are connected (fourth row) to allow control at the application layer and to implement quantum algorithms e.g. Grover's, Shor's and quantum simulation algorithms applied to problems where classical computation is unpractical [62].

The physical size function of a logical qubit and of a logical qubyte, conventionally defined above to calculate the maximum density of the logical qubits by including interconnections, are reported in last two rows in Table I. About the footprint of classical control circuits, the Steane code layout carries some unused area, which reaches about 23 % in the logical qubit and 43 % in the qubyte mask, that could be employed for classical electronics and low-level interconnections.

The surface code layout is more compact and so much more challenging to this regard, since even elementary circuits cannot fit into the small distance between physical qubits (the physical qubit pitch is of the same size order of the qubit itself as well as of a typical CMOS transistor). While increasing this feature size would be detrimental for the surface code operation and performance, a viable alternative could be to conceive a quantum processor architecture based on separated blocks of physical qubits interleaved by blocks of classical control circuits, as proposed in Ref. [112]. Each block, corresponding to one or few logical qubits, is connected to the neighbors by long-range quantum communication channels. Part of these blocks could be also reserved as ancillae factories to the high-throughput generation and purification of high-fidelity ancillae and cat states, which accounts for most of the resources involved in Shor's algorithm.

However, the operation of two-qubit gates within this topology and the effectiveness of such long-range couplers is still to be carefully evaluated in terms of fidelity and of required transfer bandwidth: the coherent transfer of a realistic logical qubit ($d \approx$ 20-30) means the coherent transfer of $d^2 \approx$ 400-900 physical qubits over few tens of μm on such a single track.

Anyway, the footprint of integrating classical control circuits obviously depends on the complexity of the functionality that is required within the quantum layer and this is a key point that the next-generation quantum engineer will have to deal with. This falls well beyond the scope of this review: we only note that, although we did not estimate the footprint of control circuits, we obtain an estimate of the *maximum* density i.e. a higher bound of quantum information. We will show how the evaluation of such figure of merit across different technology nodes, together with practical considerations about the hybrid qubit architecture, can give useful indications to the development of quantum computers.

## II. TECHNOLOGY-DEPENDENT SCALING IN THE QUANTUM REALM

We turn now to the evaluation of the physical sizes of such integrated quantum circuits in various technology nodes from 7 nm to 65 nm. The mask layouts shown in Figure 1 represent devices involving hybrid properties that cannot be univocally ascribed to any existing classical device, which rely on a fabrication precision currently unaddressed. Table II reports the minimum feature sizes of the above quantum devices as a function of the technology node, starting from 65 nm down to the 7 nm node by following the miniaturization from left to right. We associate the contacted gate pitch $\Delta_G$ to double the gate length of a state-of-the-art MOSFET. Forecasts of the forthcoming nodes at 10 nm and 7 nm were taken from the ITRS estimation of the uncontacted polysilicon pitch in flash memory. The definition of vias sets stringent requirements related to the alignment to higher level interconnects. Therefore, the interconnect pitch $\Delta_{IC}$ is estimated from the pitch of the first level of metal interconnects in microprocessor units (MPU).

The geometrical shape of quantum dots is maintained across the different technology nodes by applying an equivalent scaling of the width of the silicon islands $w_{Si}$ and of the metal lines $\Delta_G$, corresponding to the lateral sizes of the quantum dots. We set $l_{HDD} = \Delta_{IC}$ to provide reasonably wide doped regions for SET ohmic contacts and electron reservoirs. Finally, the undoped buffer length must be much larger than the lateral straggling range due to the dopant implantation process. Low energy implantation results in shallow distribution of dopants with a reduced lateral straggling of about 5-10 nm [69], pointing to the reasonable condition $l_{BU}$ > 20 nm. However, a large $l_{BU}$ size is required at SET 1 in Figure 1 to accommodate all the adjacent gates and interconnections [60]. As a consequence, we set $l_{BU} = 2\Delta_{IC}$ to allow a realistic routing of the metal wires. We note that in all the technology nodes this size is significantly larger than the dopant scattering length. Therefore, electrically active dopants in the active region of the device are excluded.

Table III reports the physical sizes of the three building blocks used for the Steane code, of the logical qubit and the qubyte at different technology nodes. Figure 3a compares the scaling trend of the area of a typical classical reference device namely SRAM 6T cell with the physical qubits namely the D module and the single hybrid qubit for the Steane and the surface codes respectively. As we are interested in a possible link between classical and quantum technologies, we focus on the product of the two critical pitches, namely the connected gate and interconnect pitch, as a relevant figure of merit to compare size scaling of classical and quantum devices. The area of the 6T-SRAM cell scales according to Moore's law with an area of about 17 $\Delta_G \cdot \Delta_{IC}$. On the other way around, the quantum D module requires from 80 to 90 times $\Delta_G \cdot \Delta_{IC}$, corresponding to roughly 5 times the area of SRAM. As a result, the physical qubit area follows a characteristic scaling rule driven by its specific design that, however, turns out to be primarily dependent on the parameter $\Delta_G \cdot \Delta_{IC}$ and has consequently many analogies with the classical device scaling law. For example, the change of the slope below 32 nm is mainly due to the slowdown of the transistor gate length reduction.

In the case of the Steane code, the resulting density of quantum information is defined $\delta_{QI} = 8A_{QB}^{-1}$ in units of logical qubits per unit surface, and it is plotted in Figure 3b as a function of the technology node. In order to express quantum computational power, the trend line refers to the technology node. As we are interested in determining the maximum quantum computational power storable in a solid-state manufacturable chip, we link the maximum amount of quantum information to the technology node. We anticipate that considerations on the operation time obey to additional constraints, which exclude the largest technology nodes from any practical applications. The quantum information density for the Steane code is of the order of Mqubits/cm$^2$, but it decreases by two orders of magnitude if one shifts to concatenated codes. Figure 3b includes the physical footprint of logical qubits encoded by surface codes with several code distance values *d* for a comparison with Steane code and recursive coding, giving an intermediate value. Such *d* values correspond to the minimum resources needed to apply Shor's factorization algorithm to 128, 1024 and 8192 bit integer numbers. This dataset covers a wide range of realistic problems including decrypting RSA systems that are actually considered safe with present and near-mid future technology of classical computing.

As noted above, the trend line reflects the change of the slope at the 32 nm node of the Moore's law, mainly associated with the change of the scaling law of the gate length with respect to the technology node. The two facts are only indirectly related, as the slowdown in the Moore's law depends on the effort capability of semiconductor industry and not on some intrinsic physical constraint. As a matter of fact, the introduction of new technologies and materials (strained silicon channel and high-k gate stack for example) has been even more important than bare geometrical scaling to maintain the equivalent scaling of device performances during the last two decades [70]. Moore's law steered semiconductor industry to the evolution from bulk planar transistors used in 65 nm node to Silicon On Insulator (SOI) devices up to the FinFET geometry adopted in the present 14 nm node [67]. Preliminary studies predict that new materials other than Si could be employed as the channel material and 3D integration may become a cheaper alternative to continuing 2D scaling to increment functional complexity of integrated circuits [68].

In such a varying framework, some additional issues must be considered that are specific of ultra-scaled quantum technology and may generate significant deviations from the pathway foreseen for classical CMOS electronics. The main point is related to the complex design of qubit devices, which encompasses different building blocks, such as a single charge sensor and multiple quantum dots, with a large number of critical gates and components that are all fundamental for the device operation. To this extent, process variability issues are expected to be dramatically increasing, with respect to classical electronics, in view of large scale implementation of CMOS qubits. As a consequence, the development of solid design tolerant to process variability must have a key role in the development of such devices. The device mask could be significantly made simpler by introducing an emerging single-charge sensing technique based on rf-reflectometry where the charge sensor is fully integrated with the QD gate control as proposed for the surface code implementation [71, 72]. With this solution, the two single electron transistors in the D module would be unnecessary for charge sensing as well as one of the electron reservoirs, with beneficial outcomes in terms of device area and complexity (the number of critical gates would reduce by a factor of 2). Another advantage would be much faster quantum state read-out, without the slow electron tunneling time limiting the procedure at μs timescales. On the other side, such technique will require intensive use of rf signals, with the potential drawbacks of a more challenging classical control circuitry and pronounced cross-talk. Although a large part of the technological challenges towards large-scale quantum computing are shared with the development of end-of-roadmap classical technology nodes, realistic qubit implementations, while being compatible with classical electronics manufacturability limits, will also presumably come to terms with the best compromise between stringent design rules imposed at the quantum level and limited cross-talk arising from classical control circuitry.

# III. OPERATION TIME CONSTRAINTS AND SOURCES OF DECOHERENCE

The figures of merit of a physical qubit, such as the energy spectrum of the quantum dot and the intensity of the inter-dot effective interactions, are influenced by the different size of the devices achievable at the different technology nodes [73]. The vertical constraint on the silicon thickness ($t_{Si} << w_{Si}$) is related to the transverse size $w_{Si}$ and it may contribute to valley splitting so that a controlled valley states filling is obtained [43, 52] as preferable for the three spin exchange-only qubit [44]. Notably, such requirement is implicitly granted by Ultra-Thin Body SOI technology to be employed for nodes beyond 14 nm. Valley splitting can be further increased up to ~ 1 meV by applying vertical electric field [28, 74]. It is worth mentioning that the complementary realization of the CMOS qubits by hole spins would avoid such issues [47], while a hypothetical realization by donor quantum dots [29, 30] would natively induce an adequate valley splitting [31].

Virtual qubit manipulation is mediated by the effective exchange interaction $J = t^2/\varepsilon$,, where $t$ is the tunnel coupling between the energy levels in the two quantum dots and $\varepsilon$ is the system detuning [38, 44]. During qubit operations, $t$ and $\varepsilon$ are regulated by the inter-dot electrostatic barrier and by the electrochemical potential of the quantum dots respectively, that are controlled through gate electrodes. Virtual gate operation frequency is therefore strongly dependent on the physical size of the quantum dots, since it is directly proportional to the maximum exchange interaction:

$$J_{max} = \frac{t^2}{\Delta E_{ST}} \,, \tag{2}$$

where $\Delta E_{ST}$ is the singlet-triplet splitting [45].

Tunnel coupling has an exponential dependence on the inter-dot distance, that in our case is equal to the metal gate pitch $\Delta_G$ [45]. $\Delta E_{ST}$ also depends on size and geometry of the quantum dots, though an additional fine tuning is possible due to a weak dependence on vertical electric field [75]. As a result, faster operations would be possible in principle at the ultimate technology nodes, with advantages in terms of fidelity. However, an upper bound to operational speed is set by the tunnel coupling being smaller than the single particle level spacing and the singlet-triplet energy splitting [43]. In order to identify the time operation window compatible with physical constraints as a function of the technology node, we consider the three following requirements: *adiabaticity*, *coherence* and *node-dependent minimum operation time*. Firstly, adiabaticity means that tunnel coupling $t$ must be lower than half the singlet-triplet splitting $t < \Delta E_{ST}/2$ [43]. Since the operation time is $T_{op} \sim h/J_{max}$ where $h$ is the Planck constant and $J_{max}$ is the maximum effective exchange interaction as defined in Equation 2, then such condition reads $T_{op} > \frac{4h}{\Delta E_{ST}}$.
Secondly, coherent qubit manipulation is obtained provided that the logic gate operation time is much smaller than the qubit dephasing time: $\frac{T_{op}}{T_2^*} < \eta$ where $\eta$ is the error threshold for QEC [61]. Thirdly, logical operation time has also a lower bound imposed by Eq. 2 which is dominated by the exponential dependence of the tunnel coupling t on the technological node [60]. For example, for a π/8 rotation gate, operation time is the following [39]:

$$T_{\pi/8} = 3.598 \, \frac{h}{J_{max}} = 3.598 \, \frac{h \, \Delta E_{ST}}{t^2} \qquad (3)$$

In Figure 4 we summarize the constraints related to the operating frequency of Si exchange-only qubits for realistic devices, including the technological limit of manufacturability at 3 nm (orange line). This limit was discussed and quantified by Kelly in 2011 by considering a number of arguments based on the intrinsic variability of top-down fabrication processes, on the unwanted electron tunneling and on the parasitic interferences, leading to a limit on the manufacturability at around 3 nm [2]. Furthermore, we highlight the physical bounds imposed by the requirements of adiabaticity, coherence and node-dependent minimum operation time, which hamper quantum computing in the blue, red and grey area respectively. The realistic operation range covered by white areas is centered at 10 GHz and, depending on the intensity of the exchange interaction and $\eta$, it spans one decade towards both lower and higher frequency, resulting therefore of about 1-100 GHz. It is a favorable circumstance that, as shown in the Supplementary Information and discussed later, such range covers the working frequency of most classical few qubit control circuits.

The main sources of decoherence are hyperfine interaction, electrostatic noise and electron interaction with optical phonons [43, 115]. The first experimental demonstration of the exchange-only double QD qubit showed $T_2^*$ of 10 ns with operation time of the order of 100 ps with Si-SiGe QD. More in detail, 5.2 GHz X-rotations and 11.5 GHz Z-rotations were coherently driven with fidelity of 85% and 94% respectively [37]. Preliminary results were recently demonstrated for two-qubit gates in an analogous system [109]. Notably, promising results were obtained for single spin [36] and singlet-triplet [35] qubits in the same material, indicating significant improvement when spin-less material is employed to reduce spin decoherence. Indeed, hyperfine interaction and magnetic field fluctuations, that are mainly responsible for decoherence in III-V semiconductors due to the high density of spin-carrying nuclei, are effectively inhibited by utilizing purified $^{28}$Si [22, 36, 76, 77]. Alternatively, hole spin qubits could be considered in the future [46–48] to reduce hyperfine interaction. Another significant noise component arises from the interface in the case of Si-$SiO_2$ QDs in particular. In fact, interface and oxide traps act as charge fluctuators that induce dephasing due to slow variations of QD potential and high frequency random telegraph noise [78]. Such consideration is even more topical for etched SOI MOS nanostructures such those considered above, which confine electrons with a small number of electrostatic gates [79]. On the one hand, etched Si nanostructures provide enhanced electron confinement, leading to increased charging energy with respect to planar electrostatically-defined QD, as well as large orbital and valley splitting [52]. On the other hand, increased surface to volume ratio will lead to a major impact of interface-related noise with respect to spin noise especially at the most aggressive technology nodes. In this regard, adequate device post-processing is effective in leading to very low density of defects of ~ $3 \cdot 10^{10}$ cm$^{-2}$ [80], corresponding to an average distance between defects of 60 nm. Furthermore, large sweet spots can be normally recognized in the energy diagram of the exchange-only qubit where robust qubit rotations can be performed with small impact from random fluctuations of the QD potential [37, 43]. We also note that scaling down to the most extreme technology nodes, a leading role in decoherence mechanism will be taken by electron-phonon interaction instead of charge noise [81]. Such phenomena must be effectively suppressed by limiting the material disorder and by cooling the system at temperatures of the order of 10 mK that are actually at hand in state of the art dilution refrigerators [28, 36, 77]. As a final remark, the reduction of disorder and of charge noise are apparently among the main challenges to reach the ideal coherence time of ~ 1 μs [38] enabling large scale quantum computing within the stringent limits represented in Figure 4.

# IV. SCALING UP QUANTUM TECHNOLOGY: QUANTUM COMMUNICATION AND ERROR CORRECTION

We now turn to quantum communication, which becomes a fundamental aspect in some large scale fault-tolerant QIP architectures. The Steane code is compatible with coherent transfer of quantum information through different methods, namely the SWAP chain protocol, the Coherent Transfer by Adiabatic Passage (CTAP) and the qubit shuttle. The first method is based on the sequential repetition of SWAP gates between adjacent qubits. From previous estimates, for a 40 nm inter-dot distance, quantum information transfer between two adjacent data qubits (at a distance of 1 μm) can be carried out in 71.2 ns [60]. The SWAP operation speed is directly proportional to the maximum exchange interaction, that in turn has an exponential dependence on the inverse of the inter-dot distance $\Delta_G$ [39]. CTAP is a viable alternative for long distances, since the time required for quantum communication has a sub-linear dependence on the distance between qubits [82, 83]. CTAP has an additional advantage over the SWAP chain: initial loading of qubits with three electrons is not required. Qubit shuttle, finally, has been proposed for quantum communication in large scale single spin qubit architectures [84]. According to the latter, quantum information is coherently transferred by moving the potential well which confines the qubit along the communication channel.

The only condition is determined by adiabatic motion i.e. $mv^2 \ll \Delta_G$ where $m$ is the electron mass and $v$ is the speed of the flying qubit. Such requirement corresponds to a favorable upper bound of $v \ll 10^4$ m/s, which is limited mainly by the speed of the classical control circuitry. Both SWAP chain and CTAP protocols cannot be used for the simultaneous transfer of many qubits along the same quantum channel, but they can be employed for short range quantum communication e.g. during QEC. Conversely, qubit shuttle may be better exploited to transfer entire logical qubits over long distances, relying on the high operating speed. Alternatively, it is worth mentioning that surface code architecture would allow more relaxed requirements as they rely on neighboring qubit sites.

Teleportation may be another interesting option for massive long-range communication, as for the transfer of a whole encoded qubit to a different site. Moreover, teleportation was also proposed to inherently perform quantum gates at a distance by means of entanglement between far away qubits [84]. Such methods for quantum communication are compatible with the proposed hybrid qubit implementation. However, the final choice of the method (or the methods) to be implemented will be necessary driven by considering the effective fidelity and bandwidth offered by each method to meet the requirements of a specific algorithm implementation.

Error threshold strongly depends on the specific code definition, but also on the complexity of the routines for quantum error detection and correction, involving several logic gate sequences and intensive transfer of qubits. Typical error thresholds range between $10^{-6}$ and $10^{-3}$ errors per logic gate, depending on the code properties, the qubit arrangement and the physical implementation. An extensive review of quantum codes and QEC techniques can be found in Ref. [61]. The first experimental demonstration of a double quantum dot exchange-only qubit reached a fidelity of the order of ~90% [37], which is insufficient for QEC purposes. However, theoretical studies indicate that much lower error rates, approaching $10^{-4}$, can be achieved by improving device quality and by optimizing qubit parameters and control sequences [43, 117]. Analogous development will be reasonably required for two-qubit gates to obtain fault-tolerant fidelity [45,109,116]. In order to discuss the worst case applied to a CMOS implementation already discussed in literature, we consider both the Steane code and the surface code, which are potentially capable to deal with such specifications. The 7-qubit Steane code constitutes the fundamental building block for advanced coding techniques, such as recursive coding (which leads to an improvement of the gate fidelity on encoded qubits) and topological color codes (which lead to error thresholds of the order of $10^{-3}$) [85, 86]. Therefore, it helps to define a benchmark to estimate the minimum physical resources required to implement QEC with the exchange-only Si-QD qubit architecture. Furthermore, the area of a logical qubit for recursive coding is of the order of $A_D \cdot (A_{qb}/ A_D)^2$ as the Steane code logical qubits are - in this case - employed as intermediate building blocks to implement a higher level doubly-encoded qubit. Circuit complexity and computational time will scale similarly. It is worth mentioning that among alternative QEC options, recursive coding and topological codes may lead to lower error rates on the encoded qubits, thus relaxing the constraints of the operation time [15–17], with the payoff that the quantum information density can be drastically reduced as shown in Figure 3b.

Besides the protection against bit-flip and phase errors, another important aspect consists of the mitigation of the leakage errors, of charge noise disturbance and electron loss.

Quantum error correction only applies to errors within the logical subspace, therefore additional correction is needed to correct leakage errors. For surface code architecture, Preskill [110] described a gate sequence that detects leakage errors of a data qubit (D) by an ancilla qubit (A), only needed a the leakage detection unit (LDU). The method is based on measurements to detect if leakage has occurred, so the LDU inverts A if D has not leaked, while A remains unchanged if leakage has occurred. Later Mehl, Bluhm and Divincenzo [87] demonstrated a similar method called leakage reduction unit (LRU) for which the measurement process is not necessary to correct for leakage. In the case of leakage events, the definitions of D and A are then interchanged after the LRU. For Steane-code, a T gate network has been proposed to detect leakage errors (spin flips) and to replace them with errors within the logical subspace. In general, within the framework of a universal set of quantum logic ports, electron loss, similarly to spin flips, are mitigated by teleportation based gates for which the data qubit is replaced with a new ancillary qubit [84].

The effects of the charge noise have been mitigated significantly by tuning the qubit energy dispersion which is a function of the detuning between the two quantum dots [111].

Blue lines in Figure 4 indicate the physical bounds imposed by three representative values of error threshold spanning from $10^{-4}$ (reachable by Steane code [113]) up to $10^{-3}$ (concatenated and color codes [85,86]) and $10^{-2}$ (surface codes [18]) highlighting the importance of the coding technique in view of scalable quantum computing. To this purpose, Steane code approach features lower error threshold but it likely has an easier integration with the classical control circuitry. Alternatively, surface code based implementation is simpler at the quantum layer and more robust against errors at the expense of very challenging demands in classical circuitry realization. In any case, a careful analysis will still be mandatory to individuate the best coding technique for a given implementation of QEC, compatibly with the nominal fidelity of logic gates and the time load required for quantum communication during QEC operations in CMOS technology.

Looking for a fair estimation of the physical resources for quantum computation, most of the space and time resources needed to run a quantum algorithm (e.g. Shor's factorization) is generally devoted to ancillae preparation for non-Clifford gates, such as phase gates, or Toffoli gates [18]. Purification up to reasonably low error rates requires very large logical qubits, including few thousands of physical qubits. Surface codes with a minimum code distance $d = 23$ is necessary to deal with computational problems of practical interest such as the factorization of a 128 bit number [18]. The code distance is the minimum weight of a logical operator i.e. the minimum number of physical bit flips to define a logical qubit. Lattice surgery technique allows universal and scalable computation within 2D surface codes with physical resources of $\approx 8d^2$ qubits per logical qubit [18].

Notably, Table IV shows that even at the largest scale problem here considered (8192 bit factorization) the quantum computer size could be limited to few mm, which is compatible with commercial microelectronic packaging on the one hand and with the use of state-of-the-art cryostats for efficient cooling on the other hand. With regard to the execution time, we also mention that it could be further reduced either by choosing optimized versions of the algorithm [18] or by improving the 100 ns readout time set in our calculation e.g. by means of the faster rf-reflectometry readout scheme mentioned above [71,72].

# V. SCALABLE CLASSICAL CMOS CONTROL ELECTRONICS FOR FAULT- TOLERANT OPERATIONS

In double quantum dot qubit, similarly to other embodiments, controlling and interrogating virtual qubits in a loop for quantum error correction, involves the generation of nanosecond scale pulses of amplitude-modulated voltage or current. Pulses are controlled in amplitude with resolution of at least 10 bits and duration of several tens of nanoseconds, with a time resolution better than 10 ns. Since the generation of these signals is done independently and in parallel for each virtual qubit to assess dynamical decoupling and compensation schemes on related virtual gates it is necessary to implement concurrent quantum error correction loops, each connected to a single virtual qubit, similar to refresh operation of a dynamic random access memory (DRAM). As a classical electronic interface operated at room temperature involves problems such as linearity of interconnections with the number of qubits, linearly scaled thermal flux to the quantum system, and power dissipated to control each qubit before being attenuated of several orders of magnitude (often 60-100 dB), cryogenic multiplexing in CMOS is being developed [88–90]. A generic implementation of a fault-tolerant loop is shown in Figure 5a. The performance observed at cryogenic temperature of the control electronics suggests that three types of multiplexing can be used based on time-division multiple access (TDMA), frequency-division multiple access (FDMA), and space-division multiple access (SDMA) to significantly reduce the number of interconnects and reduce power dissipation to the cooling power of the refrigerator. The creation of TDMA multiplexers capable of operating at mK temperatures is required, while TDMA demultiplexers and the reminder of the loops can already operate at 1.6-4.2 K [88, 90]. The programming of the digital back-end and of the analog front-end can be done with high-speed serial lines, thus minimizing the number of interconnects connecting room temperature devices to cold circuits. In order to minimize the wiring requirements, a first layer of electronic control can be implemented as close as possible to the qubits in terms of temperature and/or physical distance, either on the same silicon substrate or via 3D integration [112, 114]. This first layer would enable the afore-mentioned multiplexing schemes and, consequently, a simplified wiring towards higher electronic control layers, which can then be operated at higher temperature with relaxed power consumption and size constraints.

Cryogenic electronics has been used in several applications to improve electronics performance, e.g. by reducing the thermal noise in readout for high-energy or nuclear physics experiments [93], or to serve in harsh environments, e.g. in space applications [94, 95]. A few attempts have been made to interface quantum devices with cryogenic electronics to reduce the wiring towards room temperature [23, 24, 26, 96–98], but those approaches have been limited to one or two quantum devices, thus not truly addressing the scalability issues of quantum computers. However, by relying on the progress of semiconductor technology, only CMOS technology can currently offer the integration of billions of transistors on a single chip, while ensuring low-power consumption, reliability and functionality down to 30 mK [99]. Beside simple cryogenic CMOS amplifiers [23, 24, 93, 100], complex cryogenic analog circuits have been implemented in CMOS, including a full 400 MHz transceiver operating at 173 K [95] and several 4 K analog-to-digital converters (ADC), such as successive-approximation-register (SAR) [25, 101], Sigma Delta [102] and Flash ADCs [92]. However, such devices operate at a relatively high temperature (>> 4 K) or they show poor power efficiency. The energy efficiency of cryogenic CMOS ADCs is above 1 pJ/conversion-step [25, 92, 101, 102], which is 200 times worse than state-of-the-art CMOS ADCs operating at room temperature [103], as shown in Figure 5b. This would result in a power consumption over 200 mW for a 10 bit 200 MSa/s ADC, as required in the quantum-processor controller. The large gap to the state-of-the-art can be attributed to the use of older technologies (feature size > 0.35 μm) and/or to the adoption of conservative circuit topologies designed to be robust to cryogenic non-idealities even when accurate device models are unavailable. State-of-the-art performance at 4 K can be achieved by implementing aggressive circuit techniques, such as digitally-assisted analog blocks and time-domain signal processing, in technology nodes beyond 40 nm CMOS and by developing relative accurate device models. Such approach, combined with an extensive multiplexing strategy, may allow meeting the objective of power dissipation in the order of 1 mW per qubit channel, thus enabling the operation of thousands of cryogenic concurrent fault-tolerant loops in existing refrigerators with cooling power in the order of 1 W at 4 K. The adoption of ultra-scaled CMOS technologies combined with the increase of the carrier mobility at 4 K may allow the operation at few GHz of amplifiers and down- and up-converters. As last remark, we observe that the operation frequency of the cryogenic electronics is compatible with the timescale of the silicon qubits discussed above (see Supplementary Material).

## VI. THE FUTURE OF CMOS QUANTUM INFORMATION PROCESSING

Many of the advances described above for quantum information processing in silicon have been mirrored, either earlier or later, by similar advances in III-V semiconductors and superconductors. The latter offer the advantage that they exploit co-integration of the control electronics and more relaxed size constraints respectively. The shorter coherence time and the large size provide also comparable disadvantages. Silicon platform is perhaps further ahead than III-V and superconductive qubits in areas where scalability and nanometric size can be used, but are behind in the control of either the electron or hole wavefunction. Furthermore, the qubit operation timescale is compatible with the GHz range operation frequency of CMOS cryogenic control electronics, so a fully integrated approach would be possible. The prospects for quantum information processing devices realized in silicon may be bright in the long run because of the superior coherence time and scalability properties. It is also a great advantage that $^{28}$Si lacks of spin-orbit interaction. As advances continue in modeling, manipulation, control and devices, the field of silicon CMOS qubits seems to be progressing to the point where the ultimate scaling of the transistor coincides with the most suitable embodiment for feasible massively parallel spin-based quantum information processing. Such trend may drive the semiconductor device community to the end of the roadmap not in the classical sense, but by moving to the quantum domain represented by CMOS-based quantum computers.

**Competing interests**

The authors declare no competing financial interests.

# FIGURES

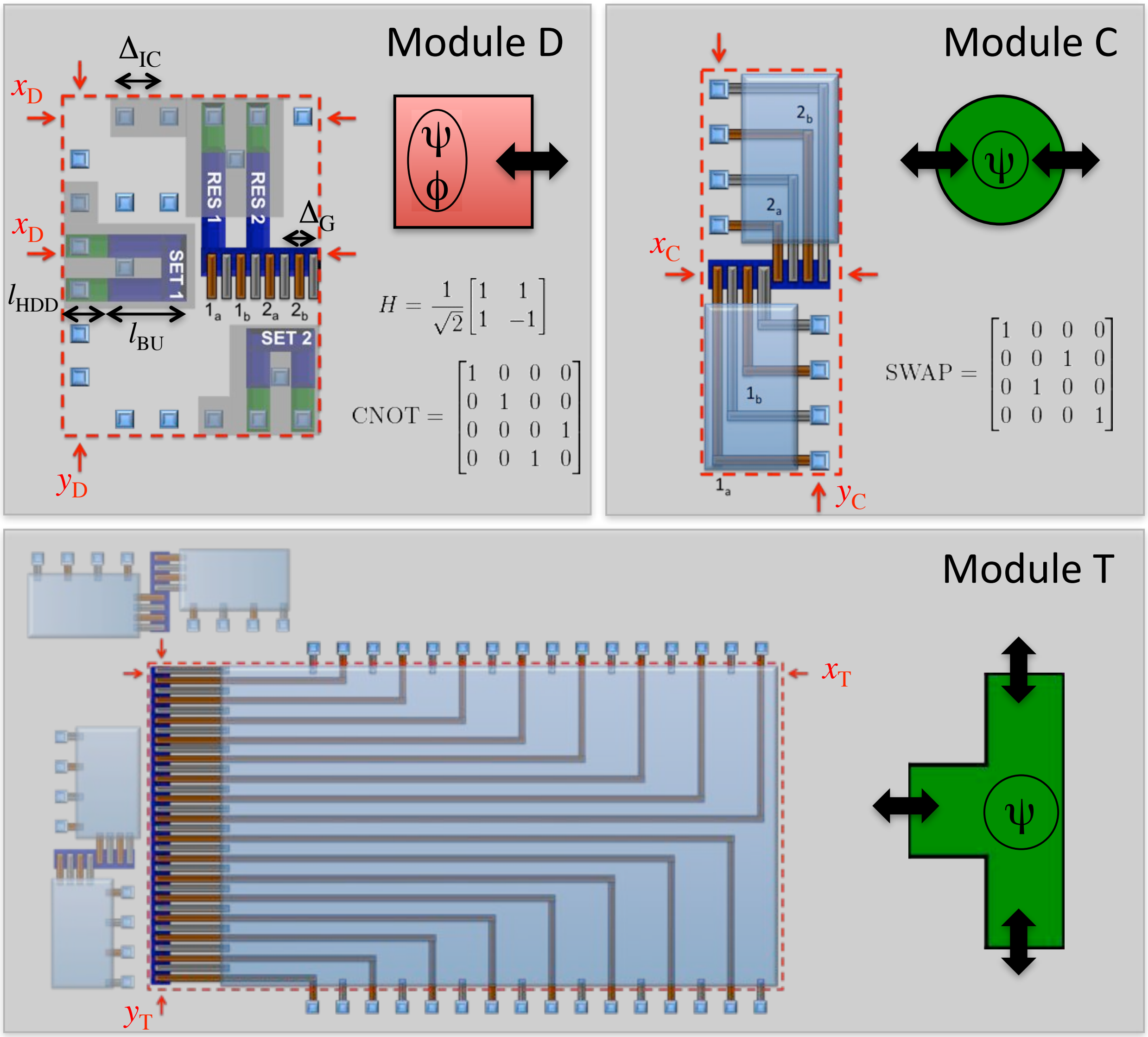

Figure 1 – The three physical building blocks constituting the virtual layer of the qubits based on exchange-only DQD qubits for the implementation of the Steane code. Schematic device mask is provided for each module together with its symbolic representation utilized in Figure 2. Greek letters denote virtual qubits storable inside the module, whereas arrows indicate the available quantum connections to neighboring virtual qubits. Top left - Data virtual qubit (D) for one- and two-qubit virtual gates e.g. the Hadamard (H) and the Controlled NOT (CNOT). Labelled plunger gates

(orange) control the chemical potentials of the four QDs defining the two qubits. Barrier gates (grey) control the inter-QD tunnel coupling. Shaded areas indicate the charge sensors (SET) and the electronic reservoirs (RES) of virtual qubits 1 and 2. Black arrows denote the critical feature sizes defined in the main text. Module sizes are calculated along the directions indicated by the red arrows. Top right - Chain module (C) for quantum communication through SWAP gates. c) Bottom - T module (T). The module supports quantum communication along the silicon nanowire direction (in this case vertically) analogously to module C. In addition, a horizontal virtual qubit chain can be connected on the left of such device to build up a T-shaped crossing with a vertical channel. Blue areas denote the space for metal interconnections between the active area over the Si nanowire and vias at the module boundary. Only plunger gates interconnections (orange) are depicted for clarity. Metal interconnections are disposed so that the space for surrounding C modules (depicted shaded on the left and top of T module) is taken into account.

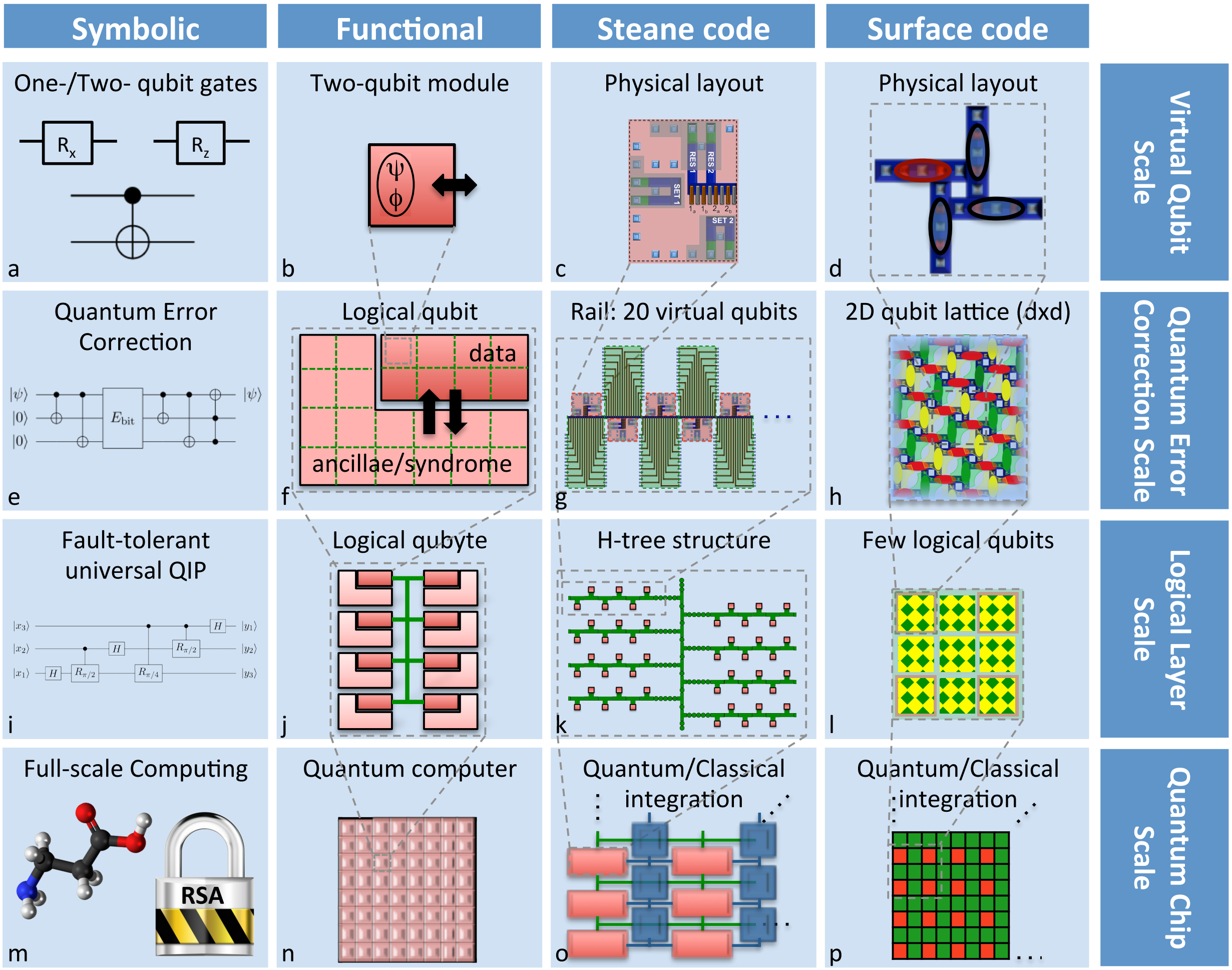

Figure 2 – Representation of QIP at different scales of integration according to the different layers of implementation.. The first column represents the different scales of integration with symbolic diagrams. The second one is a functional representation of the blocks required to execute the corresponding operations. The third column provides the physical layer implementation by the Steane code based on the modules D, C and T, while the fourth column by the surface code. Each oval indicates a double quantum dot qubit. In panel **h**, the representation of the 2D nearest neighbor array shows the virtual data qubits in red for the surface code and ancillae for the X (yellow) and Z (green) stabilizer measurements. The four rows represent four scales of integration: the virtual qubits by their physical layout, the quantum error correction scale for logical qubits, the logical layer scale and finally the quantum chip scale enabling quantum algorithms such as Shor's factoring and quantum simulations. The logical layer scale assesses fault tolerant computation from few logical qubits and it has been used to calculate the area of a single logical qubit of the Steane code by including interconnections,.

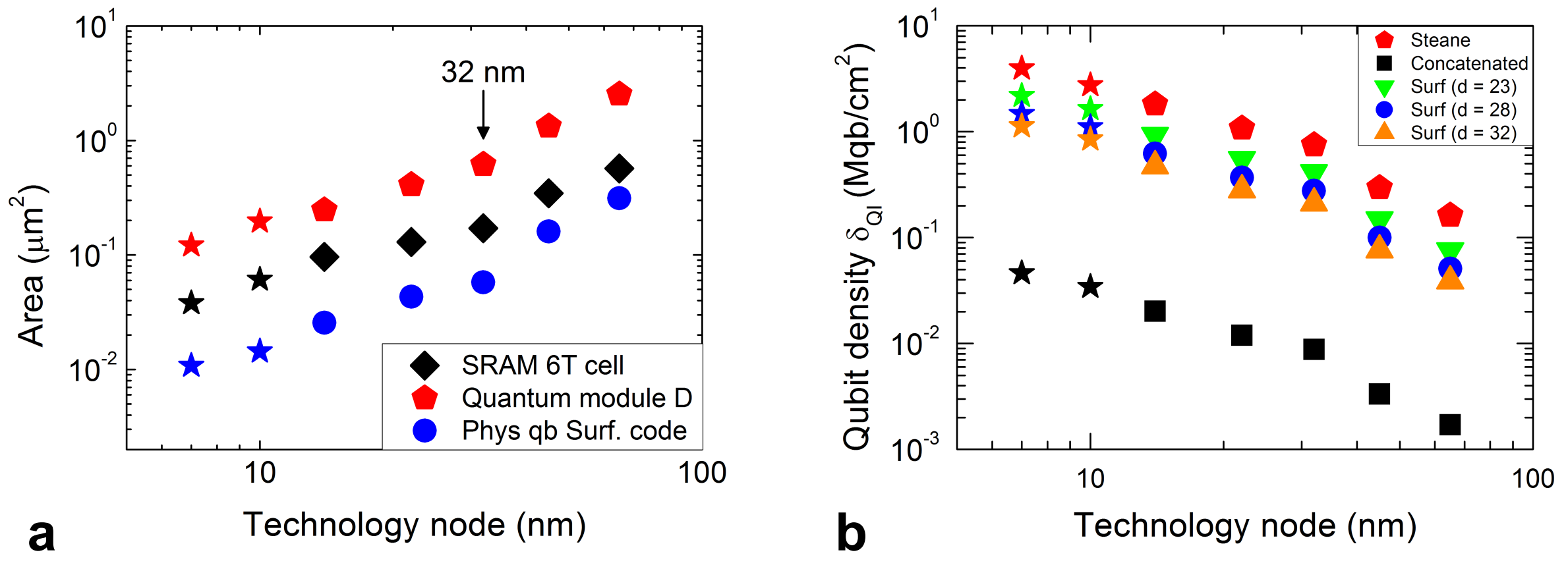

Figure 3 – The quantum information density scaling law. a) Comparison between the trend line of the area occupied by a reference SRAM classical cell and a CMOS physical qubit for both Steane (module D) and surface codes respectively. The change in the trend at 32 nm node is due to different scaling law of the gate length below 32 nm. Stars refer to the forecasts of future nodes, namely 10 nm and 7 nm, not yet available for industrial production. References for SRAM cells are the same reported in Table II. b) The quantum information density scaling law expressed as the maximum processable quantum information in terms of logical qubit number per surface unit as a function of the technology node. The concatenated code (black) is based on the Steane code [7,1,3] and lowers the quantum information density of 2 orders of magnitude with respect to the bare Steane code (red). The quantum information density is calculated for the surface codes at typical *d* values (green for *d* = 23, blue for *d* = 28, and orange for *d* = 32). The stars correspond to forecoming nodes, not yet realized and therefore subject to possible specification changes. The existing nodes and future ones (stars) are derived from IEDM and ITRS published specifications following the analysis in Table I.

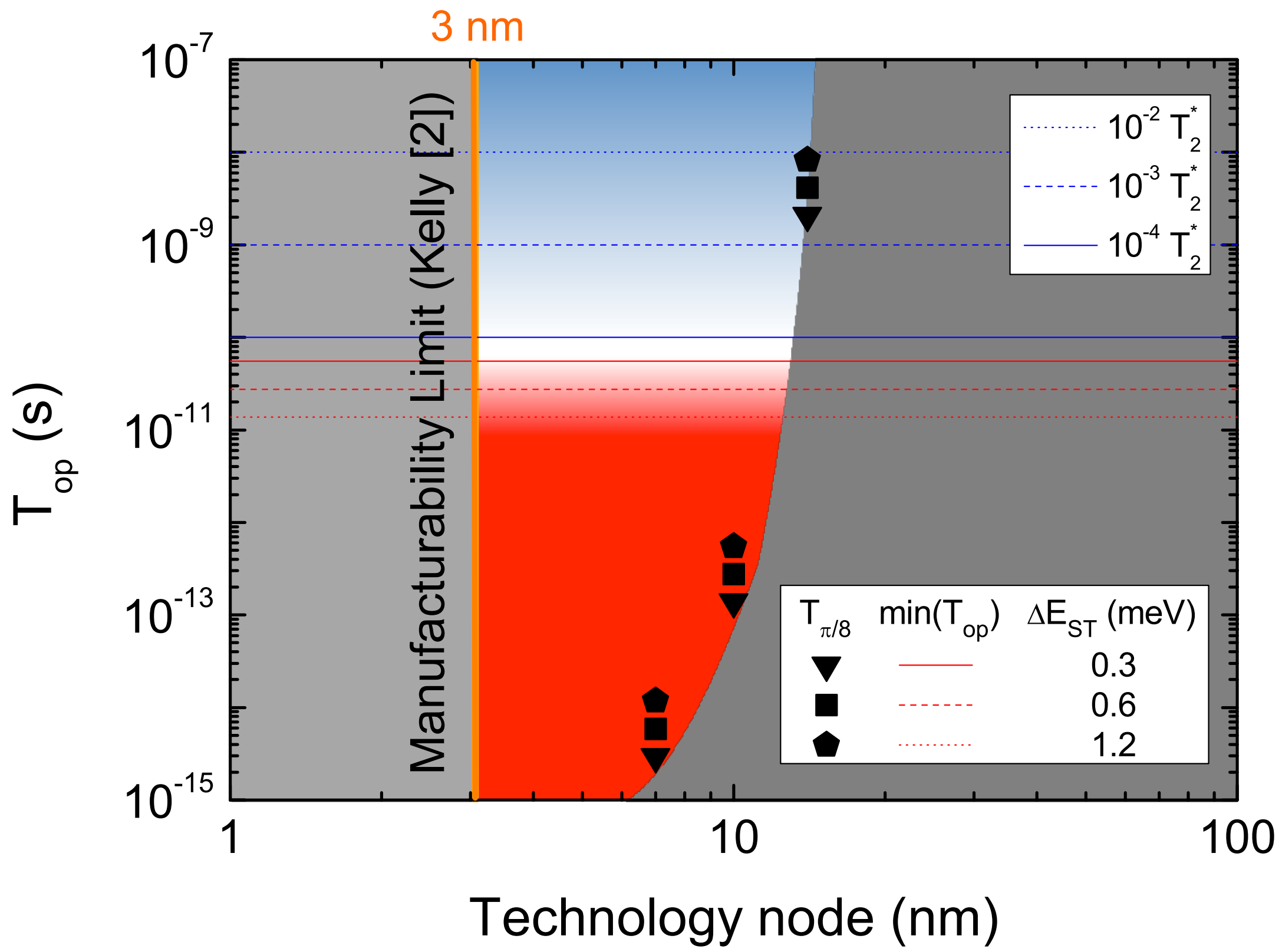

Figure 4 – Physical constraints for QIP with Si hybrid qubits. Technological limits prohibit quantum computation in the domain of grey areas due to manufacturability limits on the left side (set by the vertical orange line at 3 nm) and by the maximum effective exchange interaction on the right side (black scatters). Black data points denote the minimum operation time for a $\pi/8$ rotation that can be achieved at each technology node with the realistic values of $\Delta E_{ST}$ = 0.3 meV, 0.6 meV and 1.2 meV. Red lines indicate the minimum operation time imposed by adiabaticity at the same three values of singlet-triplet splitting of $\Delta E_{ST}$. Blue lines set the maximum operation time required by coherence requirement at three different error thresholds ($\eta$ = $10^{-4}$, $10^{-3}$ and $10^{-2}$ errors per logic gate) characteristic of available coding techniques, namely Steane code, concatenated/color codes and surface codes respectively. The ideal dephasing time $T^*_2$ = 1 μs is taken according to theoretical predictions in Ref. [37]. The white area represents the optimal regime of operation for quantum information processing with the Si qubit architecture here considered. Reasonable operating frequencies are in the order of 10 GHz, with bandwidth depending on the optimization of the QD parameters ($\Delta E_{ST}$, $t$) and of the QEC scheme compatibility with the requirements of coherence and adiabaticity.

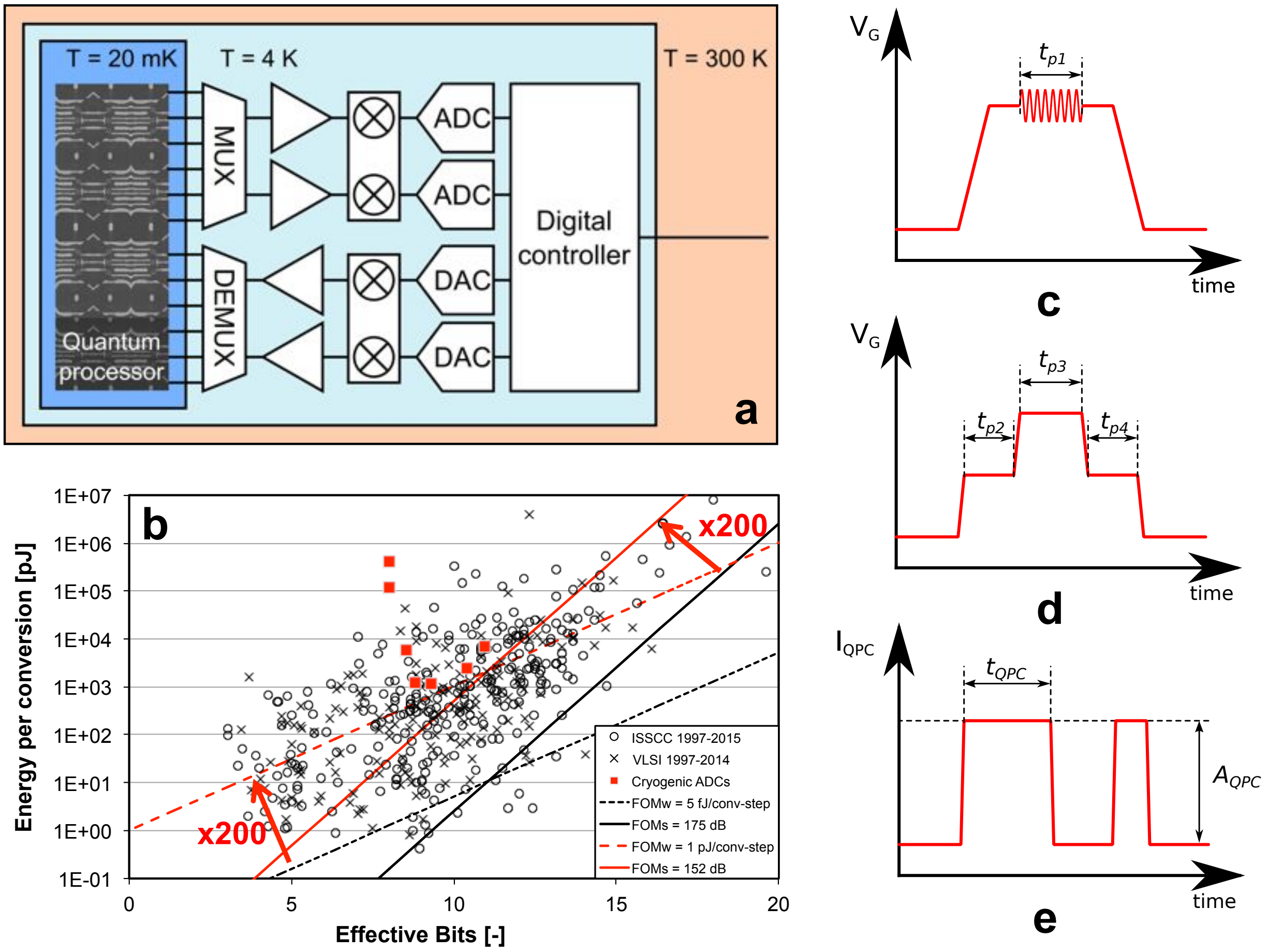

Figure 5 – Classical circuitry for qubit control and electrical signals typically used to perform operation and read out of three electron spin qubits [37, 91]. a) Generic fault-tolerant interrogation correction loop, comprising both an analog front-end and a digital back-end. b) Energy efficiency of state-of-the-art room-temperature CMOS ADCs and cryogenic CMOS ADCs [92]. The energy spent by an ADC for a single conversion, i.e. $E_C = P/f_s$ where $P$ is the power consumption and $f_s$ the sampling frequency, strongly depends on its resolution, i.e. the number of effective bits N. Consequently, ADC energy efficiency is quantified by the Schreier Figure-of-Merit (FOMs = 10 log10 [$2^N f_s/(2P)$]) for high-resolution thermal-noise-limited ADCs and by the Walden FOM (FOMw = $P/f_s/2^N$) for low-resolution ADCs. In terms of those FOMs, the energy efficiency of existing cryogenic ADCs is 200x worse than state-of-the-art room temperature ADCs (FOMs = 175 dB, FOMw = 5 fJ/conv-step), due to the use of past node technologies and the unavailability of reliable cryogenic models for CMOS devices. c,d) Voltage applied to the qubit gates to perform single-

qubit rotations; typical waveform parameters: $t_{p1}$ ~ 5-20 ns, $t_{p1}$, $t_{p2}$, $t_{p3}$ ~ 100 - 500 ps, microwave burst frequency ~ 10-12 GHz, pulse rise time (10% - 90%) ~ 80 ps. e) Typical current waveform read out at quantum point contact (QPC) to detect presence of electron in neighboring quantum dot; typical waveform parameters: $t_{QPC}$ ~ 10 ns 100 μs, $A_{QPC}$ ~ 200 pA; QPC resistance ~ 25 kΩ..

# TABLES

| Device | x | y |
|---|---|---|
| D | $x_D = \max(6\Delta_{IC}, 4\Delta_G + l_{HDD} + l_{BU} + \Delta_{IC})$ | $y_D = 8\Delta_{IC}$ |
| C | $x_C = 4\Delta_G$ | $y_C = 8\Delta_{IC} + w_{Si}$ |
| T | $x_T = 20\Delta_{IC} + l_{BU} + w_{Si}$ | $y_T = 14\Delta_G$ |
| Qubit | $x_{qb} = 20y_T$ | $y_{qb} = 2x_T - w_{Si}$ |
| Qubyte | $x_{QB} = x_C \cdot [x_T/x_C + 1] + 40y_T + w_{Si}$ | $y_{QB} = 7x_C \cdot [\max(x_T, x_D)/x_C + 1] + 2x_T + 7y_T$ |

TABLE I – Physical size of the three building blocks for QIP, of a logical qubit and of a qubyte as a function of the main critical sizes independent of the technological node. The size of the elementary components, namely the module D, C and T, are expressed by considering the physical components along the vertical and horizontal pathways between the red arrows. The sizes of logical qubit and qubyte are calculated on the basis of the physical qubit arrangement proposed in Ref. [60].

| Technology node | 65 nm | 45 nm | 32 nm | 22 nm | 14 nm | 10 nm | 7 nm |
|---|---|---|---|---|---|---|---|
| $\Delta_G$ (nm) | 140 | 100 | 60 | 52 | 40 | 30 | 26 |
| $\Delta_{IC}$ (nm) | 220 | 160 | 112 | 90 | 70 | 64 | 50 |
| $w_{Si}$ (nm) | 140 | 100 | 60 | 52 | 40 | 30 | 26 |
| $l_{BU}$ (nm) | 440 | 320 | 224 | 180 | 140 | 128 | 100 |
| $l_{HDD}$ (nm) | 220 | 160 | 112 | 90 | 70 | 64 | 50 |
| Reference | [63] | [64] | [65] | [66] | [67] | [68] | [68] |

TABLE II – Size (nm) of the critical features at different technological nodes and corresponding references from IEDM conferences and ITRS documents.

| Technology node | 65 nm | 45 nm | 32 nm | 22 nm | 14 nm | 10 nm | 7 nm |
|---|---|---|---|---|---|---|---|
| $x_D$ (nm) | 1440 | 1040 | 688 | 568 | 440 | 384 | 304 |
| $y_D$ (nm) | 1760 | 1280 | 896 | 720 | 560 | 512 | 400 |
| $x_C$ (nm) | 560 | 400 | 240 | 208 | 160 | 120 | 104 |
| $y_C$ (nm) | 1900 | 1380 | 956 | 772 | 600 | 542 | 426 |
| $x_T$ (nm) | 4980 | 3620 | 2524 | 2032 | 1580 | 1438 | 1126 |
| $y_T$ (nm) | 1960 | 1400 | 840 | 728 | 560 | 420 | 364 |
| $x_{qb}$ (nm) | 39200 | 28000 | 16800 | 14560 | 11200 | 8400 | 7280 |
| $y_{qb}$ (nm) | 9820 | 7140 | 4988 | 4012 | 3120 | 2846 | 2226 |
| $x_{QB}$ (nm) | 83580 | 60100 | 36300 | 31252 | 24040 | 18270 | 15730 |
| $y_{QB}$ (nm) | 58960 | 45040 | 29408 | 23720 | 18280 | 15896 | 12808 |
| $A_D$ (μm$^2$) | 2.5344 | 1.3312 | 0.616448 | 0.40896 | 0.2464 | 0.196608 | 0.1216 |
| $A_{qb}$ (μm$^2$) | 384.944 | 199.92 | 83.7984 | 58.41472 | 34.944 | 23.9064 | 16.20528 |
| $A_{QB}$ (μm$^2$) | 4927.877 | 2706.904 | 1067.510 | 741.297 | 439.451 | 290.420 | 201.470 |
| $\delta_{QI}$ (Mqb/cm$^2$) | 0.162 | 0.296 | 0.749 | 1.079 | 1.820 | 2.755 | 3.971 |

TABLE III – Size (*x* and *y*) of all the base (D,C,T) and composed (qubit *qb* and qubyte *QB*) devices, area A of the D module, of the logical qubit and of the qubyte, and density of quantum information $\delta_{QI}$. The latter is expressed in units of logical qubits per unit area at the different technology nodes. The estimate for the 22 nm node reported in Ref. [60] has been refined on the basis of more detailed and quantitative considerations on the arrangement of the metal gates and interconnections in the device masks.

| | | | | Area quantum computer (mm2) | | | |
|---|---|---|---|---|---|---|---|
| N bits | Data qubits | Surface code d | Physical qubits | 14nm node | 10nm node | 7nm node | Time (hours) |
| 128 | 256 | 23 | 4.21E+08 | 10.8 | 6.1 | 4.6 | 0.01 |
| 256 | 512 | 26 | 5.48E+08 | 14.0 | 7.9 | 5.9 | 0.06 |
| 512 | 1024 | 28 | 6.77E+08 | 17.3 | 9.8 | 7.3 | 0.45 |
| 1024 | 2048 | 30 | 8.25E+08 | 21.1 | 11.9 | 8.9 | 3.58 |
| 2048 | 4096 | 32 | 8.79E+08 | 22.5 | 12.7 | 9.5 | 28.63 |
| 4098 | 8196 | 33 | 1.15E+09 | 29.5 | 16.6 | 12.5 | 229.06 |
| 8192 | 16384 | 35 | 1.22E+09 | 31.1 | 17.5 | 13.2 | 1832.52 |

TABLE IV – Minimum space occupied by the resources for running Shor's prime number factoring algorithm to numbers of different magnitudes. The surface area of the entire quantum computer is estimated for the last three technological nodes according to Figure 3b. Running time is calculated on the basis of a surface code cycle time of 200 ns which is mainly limited by the measurement time [18,106].

# Supplementary Information

| Qubit type | Single qubit gate | | Multi qubit gate | | Readout | |
|---|---|---|---|---|---|---|
| | Envelope/Modulation | Frequency Width | Envelope | Frequency Width | Modulation | Frequency Width |
| Superconducting flux [1],[2] | Gaussian/PAM 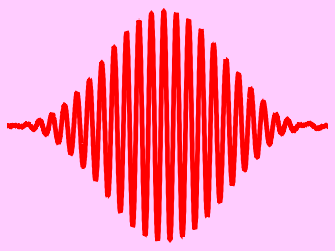 | 7-9 GHz<br>8 ns | dc pulses 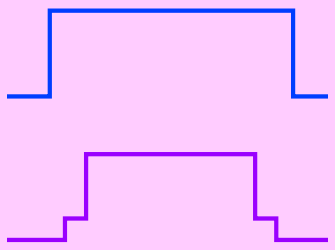 | -<br>30 ns | Amplitude 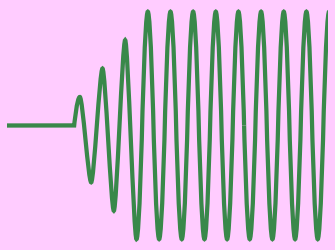 | 7-9 GHz<br>- |
| Superconducting flux [3],[4] | Gaussian/PAM 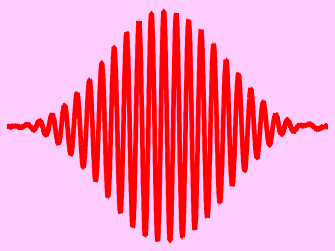 | 4-6 GHz<br>10-40 ns | dc pulses 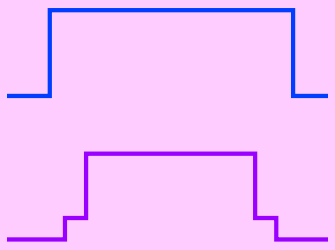 | -<br>38-45 ns | Amplitude 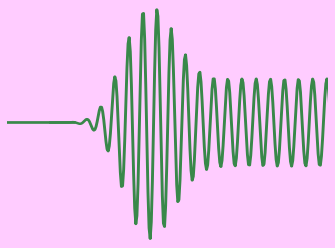 | 5-7 GHz<br>200 ns |
| Superconducting flux [5],[6] | Gaussian/PAM 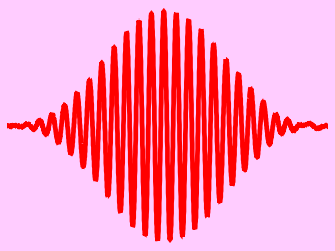 | 5-8 GHz<br>16-40 ns | Flattop Gaussian 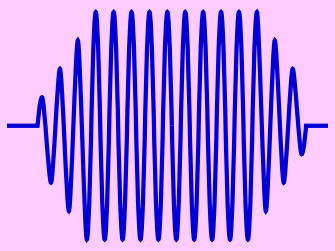 | 0.1-0.5 GHz<br>110-220 ns | Amplitude 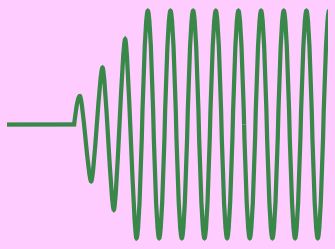 | -<br>- |
| Superconducting flux [7] | Gaussian + dc pulse 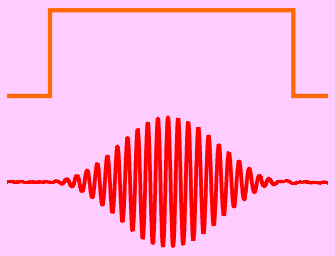 | 4-5 GHz<br>15-30 ns<br>(dc pulse:<br>0-200 ns) | - | -<br>- | Amplitude 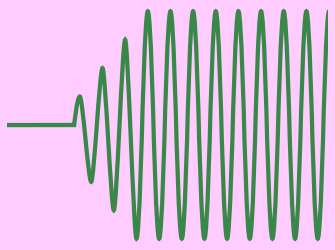 | 4-5 GHz<br>- |
| Electron spin [8],[9],[10] | Square + dc pulse 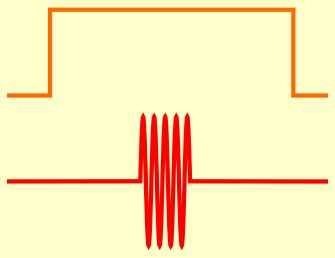 | 12.9 GHz<br>70-1200 ns<br>(dc pulse:<br>1-4 ms) | dc pulses 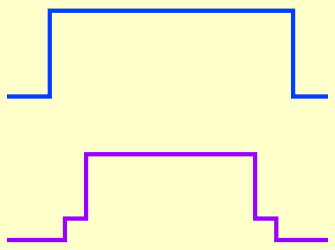 | -<br>0.5-20 ns | Amplitude 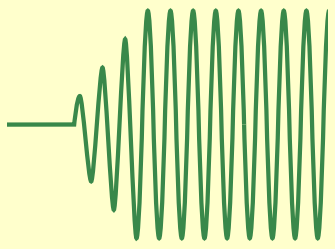 | 193.4 MHz<br>- |
| Electron spin and nuclear spin [11],[12],[13] | Square + dc 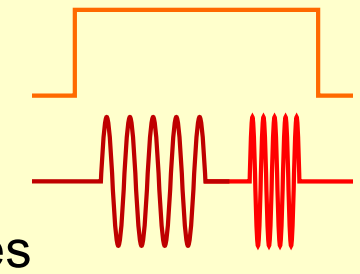 pulses | 39-49 GHz<br>(electrons)<br>14-75 MHz<br>(nuclei)<br>0-30 μs | - | -<br>- | Amplitude 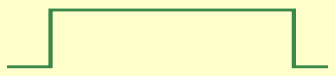 | -<br>500 μs |
| Electron spin [14],[15] | Square + dc pulses 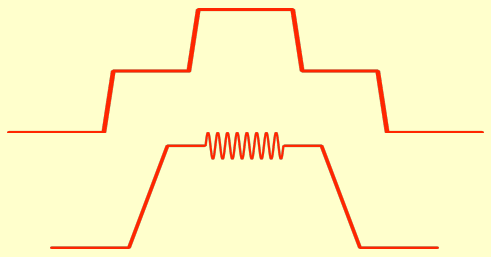 | 10–12 GHz<br>0.1-20 ns | - | -<br>- | Amplitude 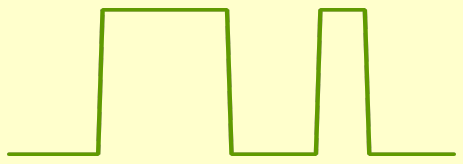 | -<br>0.01-100 ns |

Table S1

Table S1 – Overview of electric signals employed for the control of a quantum processor for several solid-state qubit technologies (superconducting-based qubits in purple rows; spin-based qubits in yellow). For each qubit technology, we report: signals to be applied to perform a single-qubit operation (typical waveform and adopted modulation in column 2; waveform parameters in column 3); signals to be applied to perform a multi-qubit operation (typical waveform, pulse envelope and adopted modulation in column 4; waveform parameters in column 5); signals to be acquired to perform qubit read-out (typical waveform and adopted modulation in column 6; waveform parameters in column 7).

Performing single-qubit and multi-qubit operations requires the generation of radio-frequency voltage pulses with Gaussian or rectangular envelope and duration ranging from few nanoseconds to few milliseconds. The carrier phase must be well-defined and its value depends on the specific operation to be performed, both when modulated using Phase-Amplitude Modulation (PAM) and when kept constant during the pulse duration. Superconducting qubits typically requires a carrier frequency ranging from 4 to 9 GHz, while spin qubits ask for frequency up to 49 GHz (for controlling electron spin) and down to 14 MHz (for controlling nucleus spin). In addition, quasi-DC voltages and/or current (not shown in Table S1) must be biased at the appropriate DC level in order to tune the qubit to the proper operation region and overcome spread in the fabrication process. Rectangular multi-level pulses with fast edges must be applied to such quasi-DC lines to temporarily move the qubit energy levels and perform single or multi-qubit operations.

Read-out can be performed (especially in superconducting qubits) by measuring the amplitude and/or phase of either the transmission coefficient between two ports of the quantum processor electrical interface or the reflection coefficient at one port. This is accomplished by generating a tone at the appropriate frequency (form hundreds of MHz to tens of GHz) and reading out the modulated transmitted or reflected tone. In spin qubits, read-out is typically performed by measuring the impedance variation of a quantum-point contact (QPC) or a quantum dot (QD). It can be done by applying a DC voltage bias to the QPC/QD and acquiring the resulting rectangular current pulse. Alternatively, the impedance can be measured by RF reflectometry, which involves sending an RF carrier (with typical frequencies up to few hundreds of MHz) to the QPC/QD and measure the variation in amplitude/phase of the reflection coefficient.

## S1 - OPERATION FREQUENCY COMPATIBILITY WITH SILICON QUBITS AND TRADEOFFS OF CRYOGENIC ELECTRONICS

Here we discuss the compatibility of the operation frequency of the cryogenic electronics with the timescale of the silicon qubits. Figure 5c-e show the electrical signals to be generated and acquired for the control of three electron spin qubits discussed in Section I. The signals are superimposed on dc voltages at multiple gates for the definition of quantum dots geometries and energy levels. Such signals are usually generated by way of an arbitrary waveform generator (AWG) with amplitude resolution of 10-16 bits and a sampling rate typically of 1 ns, capable of generating up to 20 dBm of signal strength. Nevertheless, signals of such strength and with these characteristics can be generated by ASICs (Application Specific Integrated Circuits) fabricated in CMOS technologies with feature sizes of 90 nm to 65 nm (up to 10 GHz) and 45 nm to 22 nm (up to 70 GHz). Though, the amplitude resolution may need to be relaxed to a lower number of bits. At the same time lower signal strength may result in lower power efficiency and reduced linearity. However, such reduction of specifications may be acceptable in most cases, wherever amplitude resolution may be traded for speed accuracy and higher frequencies. Since microwave frequency generation can be shared over a large number of qubits and a negligible power can be assumed for dc gate biasing, the power consumption of the classical controller is dominated by the generation and acquisition of the waveforms in Figure 5c-e. For acquisition, a power budget in the order of 10 mW would be sufficient for a room temperature read-out to meet the noise specifications (Figure 5e, 10-dB signal-to-noise ratio, 10 MHz bandwidth), and it would improve at cryogenic temperature thanks to lower noise levels. With state-of-the-art DACs already achieving energy efficiency up to 30 pJ/conversion-step at 1 GSa/s [16], a 10-bit 1-GSa/s DAC would be possible with only 30 mW, even without any optimization for the specific waveforms in Figure 5c-d. The required picosecond time resolution has already been demonstrated both in ASIC implementations [18] and in field-programmable gate array (FPGA) platforms [19], even using mature CMOS technologies (90-nm and 40-nm CMOS). The time resolution is expected to improve even further with the adoption of state-of-the-art CMOS technologies, such as 22-nm and 14-nm CMOS. With a coherence time of the order of μs and the control signal timescales in Figure 5c-e, TDMA readily allows multiplexing factors well above 10, thus demonstrating the feasibility of a power target of 1 mW/qubit-channel. With further reduction of feature sizes offered by technology CMOS nodes below 22 nm, both higher frequency and lower temperatures of operation may be achievable, thanks to the higher levels of doping in such technologies [17]. The timescale typical of qubit control electronics is summarized in the Table S1 in the Supplementary Information, where an overview of the schemes adopted for controlling qubits based on both quantum dots and superconductors is provided. The cryogenic control electronics for large scale integration therefore already demonstrated most of the properties required to fulfill qubit operation frequency range of 1-100 GHz, currently controlled with laboratory equipment (Table S1), with the prospect of ensuring a manageable thermal load in terms of power dissipation by reducing the technology node. As a result of the quadratic dependence of power dissipation from supply voltage, it will be advantageous to scale to smaller technology nodes, roughly doubling the number of channels with each technology node change. At the same time, a number of circuits, in particular multiplexers and amplifiers may be implemented to operate in sub-K with minimal or near-zero power dissipation. Such trend will enable the reduction of channels, in addition to FDMA and SDMA techniques, therefore reducing the overall channel requirements per qubit to a value below 1. This may enable further scaling of the number of qubits, with less effort in terms of power dissipation reduction.